\documentclass[10pt,conference]{IEEEtran}
\usepackage{cite}
\usepackage{pgfplots}

\pgfplotsset{compat=1.10}

\usepgfplotslibrary{fillbetween}

\usepackage{bm}
\usepackage{lipsum}
\usepackage{makeidx}
\usepackage{enumerate}
\usepackage{color}
\usepackage{cite}
\usepackage{amsmath,amsthm}   
\usepackage{amssymb}
\usepackage{nomencl}
\usepackage{multirow}
\usepackage{graphicx}
\usepackage{epstopdf}
\usepackage{threeparttable}
\usepackage{multicol}

\usepackage[mode=buildnew]{standalone}
\DeclareGraphicsExtensions{.pdf,.jpeg,.png,.jpg,.emf,.eps}
\usepackage{calc}
\usepackage{here}
\usepackage{xurl}

\usepackage{upref}
\usepackage{comment}
\usepackage{times}
\usepackage{dsfont}
\usepackage{epic,eepic}
\usepackage{rawfonts}
\usepackage[T1]{fontenc}
\usepackage{latexsym}
\usepackage{amsfonts}
\usepackage{geometry}
\usepackage[font=small,labelfont=bf]{caption}
\usepackage[font=small,labelfont=bf]{subcaption}

\usepackage{xcolor}
\usepackage{tikz,pgfplots}
\usetikzlibrary{plotmarks}
\usetikzlibrary{arrows.meta}
\usepackage{algpseudocode}
\usepackage{algorithmicx}
\usepackage[ruled,noline,linesnumbered]{algorithm2e}

\usetikzlibrary{calc}
\usetikzlibrary{intersections}
\usetikzlibrary{arrows,shapes}
\usetikzlibrary{positioning}

\newtheorem{lemma}{Lemma}

\theoremstyle{definition}

\def\R{{\bf R}}

\def\Q{{\bf Q}}
\def\I{{\bf I}}
\def\A{{\bf A}}

\def\V{{\bf V}}
\def\F{{\bf F}}

\def\G{{\bf G}}
\def\H{{\bf H}}

\def\r{{\bf r}}
\def\s{{\bf s}}

\def\x{{\bf x}}

\def\Z{{\bf Z}}

\def\Thetab{\bm{\Theta}}

\def\Psib{\bm{\Psi}}
\def\Upsilonb{\bm{\Upsilon}}

\def\tr{\operatorname{tr}}
\def\det{\operatorname{det}}
\def\diag{\operatorname{diag}}
\def\Re{\operatorname{Re}}

\def\vec{\operatorname{vec}}
\def\Exp{\operatorname{E}}

\begin{document}

\title{
MIMO Capacity Maximization with Beyond-Diagonal RIS
\thanks{This paper received support from the Smart Networks and Services Joint Undertaking (SNS JU) under the European Union’s Horizon Europe research and innovation programme within 6G-SENSES project (Grant Agreement No 101139282). The work of I. Santamaria was also partly supported under grant PID2022-137099NB-C43 (MADDIE) funded by MCIN/AEI /10.13039/501100011033.}}

\author{%
    \IEEEauthorblockN{Ignacio Santamaria$^1$, Mohammad Soleymani$^2$, Eduard Jorswieck$^3$, and Jes{\'u}s Guti{\'e}rrez$^4$}
    \vspace{0.20cm}
    \IEEEauthorblockA{$^1$Dept. of Communications Engineering, Universidad de Cantabria, Spain \\ $^2$Signal and System Theory Group, Universit{\"a}t  Paderborn, 33098 Paderborn, Germany \\ $^3$Institute for Communications Technology, Technische Universität Braunschweig, 38106 Braunschweig,
Germany \\ 
$^4$IHP - Leibniz-Institut
f{\"u}r Innovative Mikroelektronik, 15236 Frankfurt (Oder), Germany \\ 
    %Email: diego.cuevas@unican.es, carlos.beltran@unican.es, mikel.gutierrez@alumnos.unican.es, \\ i.santamaria@unican.es, vit.tucek@huawei.com, gunnar.peters@huawei.com
    }
}

\maketitle

\begin{abstract}
    This paper addresses the problem of maximizing the capacity of a multiple-input multiple-output (MIMO) link assisted by a beyond-diagonal reconfigurable intelligent surface (BD-RIS). We maximize the capacity by alternately optimizing the transmit covariance matrix, and the BD-RIS scattering matrix, which, according to network theory, should be unitary and symmetric. These constraints make the optimization of BD-RIS more challenging than that of diagonal RIS. To find a stationary point of the capacity we maximize a sequence of quadratic problems in the manifold of unitary matrices. This leads to an efficient algorithm that always improves the capacity obtained by a diagonal RIS. Through simulation examples, we study the capacity improvement provided by a passive BD-RIS architecture over the conventional RIS model in which the phase shift matrix is diagonal.
\end{abstract}


\section{Introduction}
Reconfigurable intelligent surfaces (RISs) are emerging as a new paradigm for the sixth generation (6G) of wireless communications. Composed of a large number of passive elements that can be tuned to modify their electromagnetic reflection properties, an RIS can modify the propagation channel to achieve favorable conditions for wireless communications \cite{{RenzoJSAC2020},{pan2020multicell},{ZapponeTWT2019},{ZhangTWT2019},{SoleymaniTVT2023}}. Conventional RIS models, called single-connected RIS, lead to diagonal scattering matrices. Recently, the concept of passive beyond-diagonal RIS or BD-RIS has been proposed as a generalization in which all RIS elements can be connected by variable reactances to each other, thus leading to fully connected scattering matrices which have to be unitary and symmetric  \cite{ClerckxTWC22a},\cite{ClerckxTWC22b},\cite{ClerckxTWC24a},\cite{NeriniTWC2023}. These constraints introduce new challenges in the optimization of the BD-RIS matrix for different scenarios. 

In \cite{ClerckxTWC24a},\cite{ClerckxJSAC23}, the authors consider a multi-user multiple-input single-output (MU-MISO) downlink channel assisted by a multi-sector BD-RIS in which the RIS matrices of each cell are single-connected and therefore diagonal. The optimization of multi-sector BD-RIS coefficients is performed on the complex sphere manifold. In \cite{ClerckxTWC22b}, the same authors consider more general BD-RIS models including fully-connected architectures with hybrid (reflective + transmissive) BD-RIS, and perform the optimization of the BD-RIS matrix on the Stiefel manifold. Optimization techniques for
BD-RIS with a group-connected architecture of group size two have been proposed in \cite{SoleymaniTWC2023} which apply the convex-concave procedure to convexify the unitary constraint. Closed-form fully-connected BD-RIS solutions that maximize the equivalent channel gain exist for single-input single-output (SISO) and MISO/SIMO channels \cite{SantamariaSPLetters2023, NeriniTWC2023}. The solutions in \cite{NeriniTWC2023} and \cite{SantamariaSPLetters2023} are based on two different symmetric and unitary BD-RIS matrix factorizations, but they are otherwise equivalent. More recently, an approximate solution that maximizes the sum of the equivalent channel gains when the direct channel is not blocked in an MU-MISO scenario has been described in \cite{MaoCL2024}. This suboptimal solution is based on solving a relaxed problem (without considering the unitary and symmetric constraint) and then projecting the relaxed solution onto the set of symmetric and unitary matrices, for which a closed-form solution exists. The optimization of a BD-RIS to maximize the power received by users in an MU-MISO scenario is considered in \cite{LatvaArxiv2024}. This work considers a frequency-dependent model for the BD-RIS, but the optimization is performed at a certain frequency called priority frequency. Moreover, the restriction of a lossless BD-RIS is not imposed in \cite{LatvaArxiv2024} and therefore the BD-RIS matrix is not necessarily unitary, only symmetrical.

Most of the aforementioned works consider either SISO or MISO/SIMO systems, maximizing the equivalent channel gain or the sum rate in the downlink channel \cite{SoleymaniTWC2023, NeriniTWC2023, ClerckxTWC24a, ClerckxJSAC23, ClerckxTWC22a}. In this work, we consider a multiple-input multiple-output (MIMO) link, possibly with an unblocked direct channel, and propose an algorithm to maximize its achievable rate by alternately optimizing the transmit covariance matrix and the fully-connected BD-RIS. The Takagi decomposition \cite{Takagi}, \cite{Autonne} allows us to factorize the symmetric and unitary BD-RIS matrix as the product of a unitary matrix and its transpose. This factorization is exploited in the paper to derive an optimization algorithm on the manifold of unitary matrices. The proposed algorithm provides rate improvements over a diagonal RIS that become more significant as the number of streams, the number of BD-RIS elements, or the transmitted power increases.

% \textit{Notation}: Matrices are denoted by bold-faced upper case letters, column vectors are denoted by bold-faced lower case letters, and scalars are denoted by light-faced lower case letters. The Euclidean norm is denoted by $\|v\|$, the operator norm, which is the largest singular value of a matrix, is denoted by $\|\M\|_{op}$ and $j$ denotes the imaginary unit. The superscripts $(\cdot)^{\textnormal{T}}$ and $(\cdot)^{\textnormal{H}}$ denote transpose and Hermitian conjugate, respectively. We denote by $\I_n$ the identity matrix of size $n$. ${\cal CN}(0,1)$ denotes a complex proper Gaussian distribution with zero mean and unit variance, $\x \sim {\cal CN}_{n}({\bf 0}, \R)$ denotes a complex Gaussian vector in $\mathbb{C}^n$ with zero mean and covariance matrix $\R$. For real variables we use $\x \sim {\cal N}_{n}({\bf 0}, \R)$. The Grassmann manifold $\mathbb{G}\left(1,\mathbb{C}^T\right)$ is the space of $M$-dimensional subspaces in $\mathbb{C}^T$.

%%%%%%%%%%%%%%%%%%%%%%%%%%%%%%%%%%%%%%%%%%%%%%%%%%%%%%%%%%%%%%%%%%%%%%%%%%%%%%%%%%%%%%%%%%%%%%%%%%%%%%%%%%%%%%%%%%%%%%%%%%%%%
%%%%%%%%%%%%%%%%%%%%%%%%%%%%%%%%%%%%%%%%%%%%%%%%%%%%%%%%%%%%%%%%%%%%%%%%%%%%%%%%%%%%%%%%%%%%%%%%%%%%%%%%%%%%%%%%%%%%%%%%%%%%%
 
\section{System Model and Problem Formulation}\label{sec:preliminaries}
\subsection{System Model}
We consider a MIMO link assisted by a BD-RIS in which the transmitter is equipped with $N_T$ antennas, the receiver is equipped
with $N_R$ antennas, and the RIS has $M$ elements. We assume perfect channel state information (CSI) at both the transmitter and the receiver. The equivalent $N_R \times N_T$ MIMO channel is
\begin{equation}
\H_{eq} = \H + \F \Thetab \G^H ,
\label{eq:MIMOchanneleq}
\end{equation}
where $\G \in \mathbb{C}^{N_T \times M}$ is the channel from the transmitter to the RIS, $\F \in \mathbb{C}^{N_R \times M}$ is the channel from the RIS to the receiver, $\H \in \mathbb{C}^{N_R \times N_T}$ is the MIMO direct link, and $\bm{\Theta}$ is the $M \times M $ RIS matrix. 

A fully connected reflective BD-RIS can be modeled as an $M$-port reciprocal network where each port is connected to all other ports by a reconfigurable reactance. The feasibility set for the BD-RIS scattering matrix is \cite{ClerckxTWC22a},\cite{ClerckxTWC22b}, \cite{NeriniTWC2023}
\begin{equation}
{\cal{T}}_{1}= \left\{ \Thetab \,\, | \,\,\, \Thetab^H \Thetab = \I_M, \,\, \Thetab = \Thetab^T \right\},
\label{eq:feasibilityset_BDRISpass}
\end{equation}
where the unitary constraint $\Thetab^H \Thetab = \I_M$ means that the RIS is passive lossless, whereas the symmetry constraint $\Thetab = \Thetab^T$ comes from the fact that the RIS is a reciprocal passive network for which the power losses are the same between any pair of ports regardless of the direction of propagation.

\subsection{Problem Formulation}
The transmitter sends proper Gaussian signals $\x \sim \mathcal{CN}({\bf 0}, \R_{xx})$, with $\R_{xx} = \Exp[\x\x^H] \succeq {\bf 0}$ and $\tr(\R_{xx}) \leq P$. Under these assumptions, the maximization of the capacity of a BD-RIS-assisted MIMO link can be formulated as follows:
\begin{subequations}
\begin{align}
({\cal P}_1): \,\max_{\Thetab, \R_{xx}}\,\,& \log \det \left( \I_{N_R} + \frac{1}{\sigma_n^2}\H_{eq} \R_{xx} \H_{eq} \right) \label{eq:Cap}  \\
\,\,& \tr(\R_{xx}) \leq P, \quad  \R_{xx} \succeq {\bf 0}\label{eq:TxPowerConstraint} \\
\,\,& \Thetab \in {\cal{T}}_1 \label{eq:feasibilityRIS} .
\end{align}
\end{subequations}
where $ \sigma_n^2 \I_{N_R} $ is the noise covariance matrix of the additive white Gaussian noise at the receiver, and $\H_{eq}$ is the equivalent channel in \eqref{eq:MIMOchanneleq}. Besides, \eqref{eq:TxPowerConstraint} are the transmit covariance matrix constraints. In this paper, $\log$ denotes the natural logarithm [nats]. 

%%%%%%%%%%%%%%%%%%%%%%%%%%%%%%%%%%%%%%%%%%%%%%%%%%%%%%%%%%%%%%%%%%%%%%%%%%%%%%%%%%%%%%%%%%%%%%%%%%%%%%%%%%%%%%%%%%%%%%%%%%%%%
%%%%%%%%%%%%%%%%%%%%%%%%%%%%%%%%%%%%%%%%%%%%%%%%%%%%%%%%%%%%%%%%%%%%%%%%%%%%%%%%%%%%%%%%%%%%%%%%%%%%%%%%%%%%%%%%%%%%%%%%%%%%%

\section{Capacity maximization with passive BD-RIS}
\label{sec:BDRISpass}

%The proposed method alternately optimizes the transmit covariance matrix for a fixed RIS matrix and the RIS matrix for a fixed covariance matrix. It is this second stage of the alternating optimization process that leads to a more challenging optimization problem. To solve it, we apply a minimization-maximization algorithm \cite{sun2017majorization}. The minimization step is based on a well-known concave lower bound of the capacity which leads to a convex majorization problem that can be solved by conventional techniques.

As it is usually done to solve this type of non-convex problem, we consider an alternating optimization procedure in which the covariance matrix $\R_{xx}$ is first optimized for a fixed $\Thetab$ and subsequently $\Thetab$ is optimized for a fixed $\R_{xx}$, repeating the iterations until convergence.

\subsection{Optimizing $\R_{xx}$ for a given $\Thetab$}
\label{sec:OptRxx}
For a fixed $\Thetab$, finding the optimal $\R_{xx}$ that maximizes capacity is a convex problem whose solution is $\R_{xx} = \V {\bf P} \V^H$ where $\V$ is the left eigenspace of the equivalent MIMO channel matrix $\H_{eq}$ and ${\bf P}= \diag(P_1,\ldots, P_d)$ is a diagonal matrix where $P_i$ denotes the optimal power allocated to the $i$th stream given by the water-filling strategy to satisfy the power constraint $\sum_i P_i = P$.
%%%%%%%%%%%%%%%%%%%%%%%%%%%%%%%%%%%%%%%%%%%%%%%%%%%%%%%%%%%%%%%%%%%%%%%%%%%%%%%%
\subsection{Optimizing $\Thetab$ for a given $\R_{xx}$}
\label{sec:OptTheta}
Once $\R_{xx}$ is obtained we can define $\overline{\H} = \H \V {\bf P}^{1/2}$ and $\overline{\G} = {\bf P}^{1/2} \V^H \G$. The new equivalent channel, with the $\R_{xx}$ absorbed in $\overline{\H}$ and $\overline{\G}$, is denoted as $\overline{\H}_{eq}(\Thetab) = \overline{\H} + \F \Thetab \overline{\G}^H$, where we now stress the dependency of the channel on the scattering matrix. The max-capacity problem involving only the RIS is
\begin{subequations}
\begin{align*}
({\cal P}_2): \,\max_{\Thetab \in {\cal{T}}_{1}}\,\,& \log \det \left( \I_{N_R} + \frac{1}{\sigma_n^2}\overline{\H}_{eq}(\Thetab)\overline{\H}_{eq}(\Thetab)^H \right). %\label{eq:CapRISonly} 
\end{align*}
\end{subequations}
Problem ${\cal P}_2$ is non-convex because the capacity function is non-concave over $\Thetab$. The standard technique to solve this type of problem consists of applying a minorize-majorization (MM) approach \cite{sun2017majorization} based on a concave lower bound expressed in the following lemma. This lower-bound is proved in \cite[Appendix B]{TCOMbounds16} and has been used successfully multiple times in rate optimization problems \cite{SoleymaniTVT2023,SoleymaniTSP2023,NguyenTSP2019}.

\begin{lemma}[\hspace{-0.02cm}\cite{TCOMbounds16}]
\label{Lemma:lowerbound}
Let us denote
\begin{equation*}
C(\Thetab) = \log \det \left( \I_{N_R} + \frac{1}{\sigma_n^2}\overline{\H}_{eq}(\Thetab)\overline{\H}_{eq}(\Thetab)^H \right),
\end{equation*}
given a feasible solution at iteration $t$, $\Thetab_t$, the following concave-lower bound of the capacity exists.
\begin{align*}
\label{eq:lowerbound}
    C(\Thetab) &\geq C_t + \frac{2}{\sigma_n^2}\Re  \left \{ \tr(\H_t ( \overline{\H}_{eq}(\Thetab) -\H_t)^H) \right \} \nonumber \\
    &  - \tr \left( \R_t^H \left( \overline{\H}_{eq}(\Thetab)\overline{\H}_{eq}(\Thetab)^H - \H_t \H_t^H \right) \right)
\end{align*}
where $C_t = C(\Thetab_t)$, $\H_t = \overline{\H}_{eq}(\Thetab_t)$, and $\R_t =  \frac{1}{\sigma_n^2} \I_{N_R} - \left(\sigma_n^2 \I_{N_R} + \H_t \H_t^H\right)^{-1}$. 
\end{lemma}

Let us define $\F_{t} = \R_t^{1/2} \F$ and $\Z_t = \sigma_n^2 {\H}_{t}^H - \overline{\H}^H \R_t$. After some straightforward calculations, the concave bound can be expressed more compactly as
\begin{equation}
\label{eq:lowerbound}
    C(\Thetab) \geq A_t + J(\Thetab) , 
\end{equation}
where we have defined 
\[
J(\Thetab) = 2 \Re  \left \{ \tr( \Z_t \F \Thetab \overline{\G}^H) \right \} - \|\F_{t} \Thetab \overline{\G}^H \|_F^2.
\]
The minorizer of the capacity in \eqref{eq:lowerbound} is a quadratic function of the BD-RIS matrix $\Thetab$. Therefore, the optimization problem that remains to be solved is the maximization of $J(\Thetab)$ subject to the unitary and symmetric constraints for the BD-RIS matrix. To this end, we apply Tagaki's factorization which allows us to write the optimization problem only under the constraint of unitarity. The Takagi factorization \cite{Takagi, Autonne, SantamariaSPLetters2023} proves that any unitary and symmetric complex matrix can be factored as $\Thetab = \Q \Q^T$, where $\Q$ belongs to the manifold of complex unitary matrices denoted here as ${\cal{U}}(M)$. We use this fact to formulate the following optimization problem
%The fact that any matrix $\Q \Q^T$ with $\Q$ unitary (but not necessarily symmetric) is both unitary and symmetric is trivial, so there is no loss of generality in this parametrization. We use this fact to formulate the following optimization problem
\begin{align}
\label{eq:MaximizerBDRIS}
({\cal P}_3): \max_{\Q \in {\cal{U}}(M) }\,&  J(\Q\Q^T).
\end{align}
To solve $({\cal P}_3)$ we perform the optimization on the unitary group \cite{boumal2023intromanifolds}. The tangent plane at a point $\Q \in {{\cal{U}}(M)}$ is obtained by differentiating $\Q^H \Q= \I_M$, which yields
\begin{equation*}
    \Q^H \Delta \Q + \Delta \Q^H \Q = {\bf 0},
\end{equation*}
so the tangent plane, $\Delta \Q$, is composed of all $M\times M$ matrices such that $\Q^H \Delta \Q$ is skew-Hermitian. This is a space of real dimension $M^2$. Since the cost function is not analytic, we use Wirtinger calculus taking derivatives with respect to $\Q^*$ assuming $\Q$ constant. Furthermore, from Theorem 3.4 in \cite{Hjorungnes} we have that the direction where the real function $J$ (from now on for notational simplicity we drop the dependence of $J$ on the matrix $\Q$) has the maximum rate of change is given by the complex matrix derivative with respect to $\Q^*$
\begin{equation}
    \nabla_{\Q^*} J = \left[ \A^H 
    - (\F_t^H\F_t)(\Q\Q^T)(\overline{\G}^H\overline{\G})\right] \Q^*
    \label{eq:unc_gradient}
\end{equation}
where $\A = \G_q^H\Z_t\F$. The projection of the unconstrained gradient onto the tangent space at $\Q$ is $\pi_T (\nabla_{\Q^*}  J) = \Q {\bf S}_{skew}$, where
\begin{equation}
{\bf S}_{skew} = \left((\nabla_{\Q^*} J)^H \Q - \Q^H (\nabla_{\Q^*} J)  \right)/2
\label{eq:Sskew}
\end{equation}
is skew-Hermitian. Finally, to update the unitary matrix we move along the geodesic starting at $\Q$ (the value at the current iteration) with direction $\nabla_{\Q^*} J = \Q {\bf S}_{skew}$ as 
\begin{equation}
    \Q = \Q e^{\mu {\bf S}_{skew}},
    \label{eq:geodesic}
\end{equation}
where $e^{\A}$ denotes here the matrix exponential. The learning step size, $\mu >0$, can be conveniently chosen and adapted using a line search procedure. A summary of the proposed method is shown in Algorithm \ref{alg:BDRISopt}\footnote{Matlab code can be downloaded from \url{https://github.com/IgnacioSantamaria/Code-BD-RIS-SPAWC2024}.}.
%\href{https://github.com/IgnacioSantamaria/Code-BD-RIS-SPAWC2024}{Github}}.

\begin{algorithm}[!t]
%\small
\DontPrintSemicolon
\SetAlgoVlined
\KwIn{Initial $\Q \in {{\cal{U}} (M)}$; $\H, \F, \G$, $\sigma_n^2$, $P$, $\mu$, convergence thresholds: $\epsilon_C$, $\epsilon_J$}
\KwOut{Final BD-RIS $\Thetab = \Q\Q^T$}
\While{Cap. improvement larger than $ \epsilon_C$}{
\tcc{Update $\R_{xx}$ for a given $\Thetab$ }
Obtain $\R_{xx} = \V {\bf P} \V^H$ via SVD+waterfilling\;
\tcc{Update $\Thetab$ for a given $\R_{xx}$}
Obtain $\overline{\H} = \H \V {\bf P}^{1/2}$ and $\overline{\G} = {\bf P}^{1/2} \V^H \G$\;
\While{Cap. improvement larger than $ \epsilon_C$}{
\tcc{Minorizer: $J(\Q\Q ^T)$}
  Obtain $\H_t$, $\R_t$, $\F_{t}$, $\Z_t$, and $\A$ as indicated in Sec. \ref{sec:OptTheta} \;
%The minorizer of the capacity is
%\[ 
%J = \Re  \left \{ \tr( \Z_t \F \Q\Q ^T\G_q^H) \right \} - \|\F_{t} \Q\Q^T %\G_q^H \|_F^2
%\]\;
 \tcc{${\cal{P}}_3$: Maximize $J(\Q\Q ^T)$}
 \While{$\|J(\Thetab_{k+1})-J(\Thetab_{k})\| < \epsilon_J$ }{
Calculate $\nabla_{\Q^*} J$ as \eqref{eq:unc_gradient}  \;
Find ${\bf S}_{skew} $ as  \eqref{eq:Sskew}\;
Update $\Q$ as \eqref{eq:geodesic}
}
}
}
\caption{Max. Capacity BD-RIS}
\label{alg:BDRISopt}
\end{algorithm}

%%%%%%%%%%%%%%%%%%%%%%%%%%%%%%%%%%%%%%%%%%%%%%%%%%%%%%%%%%%%%
\section{Simulation Results}\label{sec:results}

In this section, we evaluate the results of the max-capacity BD-RIS algorithm. We compare the results with a diagonal RIS whose phase shifts have been optimized to maximize capacity using the algorithm in \cite{ZhangCapacityJSAC2020} (labeled as RIS in the figures), and the solution for a BD-RIS recently proposed in \cite{MaoCL2024} (labeled as BD-RIS Low-complexity). The latter algorithm finds a suboptimal (but closed-form) solution for the maximization of the Frobenius norm of the MIMO channel and subsequently performs the projection of this solution onto the set of unitary and symmetric matrices. 
%Although not explicitly recognized in \cite{MaoCL2024}, this projection is in fact the Takagi factorization of the matrix \cite{Takagi}, \cite{SantamariaSPLetters2023}.

\paragraph{{\bf Scenario description}}
The transmitter with $N_T$ antennas and the receiver with $N_R$ antennas are located at (0,0,1.5) and (50,0,1.5), respectively, where all coordinates are in meters. The RIS location is $(d,5,5)$ where $d$ is varied along the x-axis from $d=10 [m]$ to $d=100 [m]$.The bandwidth is $20$ MHz and the system operates at 2.4 GHz. The path loss is $PL = PL_0 -\alpha 10 \log_{10} d$ where  $PL_{0} = -28$ dB is the path loss at a reference distance of $d_0 = 1$ meter and $\alpha$ is the path loss exponent. The power spectral density for the additive noise is $\sigma_n^2 =-174 + 10\log_{10}B$ dBm. The transmit power is $P= 100 $ mW. The small-scale fading for the direct MIMO link is modeled as a Rayleigh channel, and for the large-scale fading we use a path-loss exponent $\alpha_{d} = 3.75$. The path loss exponent for the channels $\G$ and $\F$ through the RIS is $\alpha = 2$ and the small-scale fading is assumed to be Rician with a Rice factor $\gamma=3$. Other parameters are taken from \cite{soleymani2022improper}.

\begin{figure}[htb]
    \centering
\includegraphics[width=.5\textwidth]{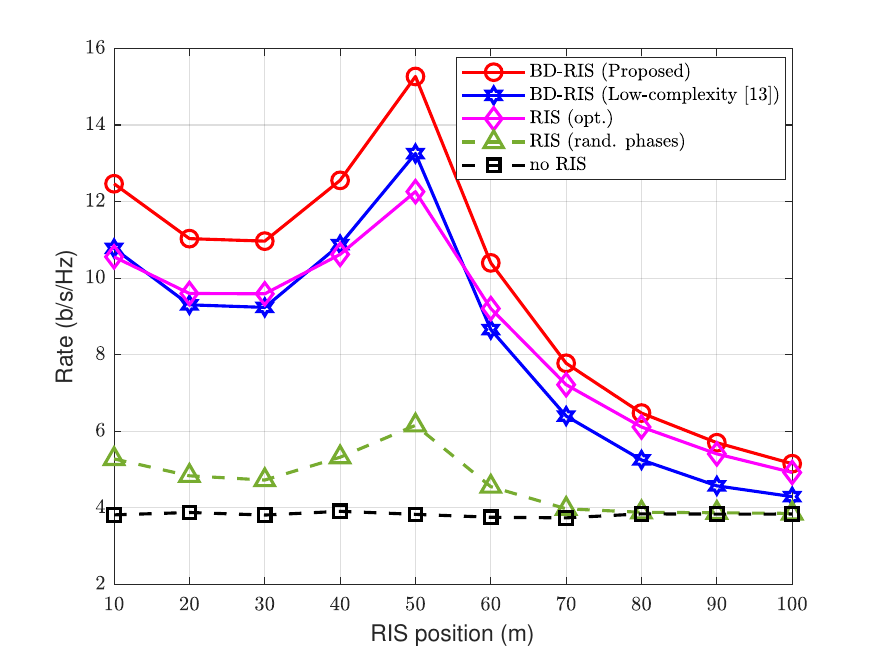}
     \caption{Achievable rate for a $4\times 4$ MIMO link assisted by an BD-RIS with $M=100$ elements.}
	\label{fig:FigScenario1SPAWC}
\end{figure}

\paragraph{{\bf Results varying the BD-RIS position}}
The first scenario considers a $4\times 4$ MIMO channel and a BD-RIS with $M=100$ elements. Fig. \ref{fig:FigScenario1SPAWC} shows the spectral efficiency in (b/s/Hz) as the BD-RIS varies its location along the x-axis from $x_{RIS} =10$ m. to $x_{RIS} = 100$ m. For each RIS position, we average the result of 100 independent channel realizations. 
The highest achievable rate is provided by the proposed BD-RIS algorithm, which always improves the results of a diagonal RIS with optimized phases, both being far superior to a diagonal RIS with random phases. As a baseline,  Fig. \ref{fig:FigScenario1SPAWC} also shows the rate without RIS. Finally, the results of the closed-form BD-RIS solution proposed in \cite{MaoCL2024} can be even worse than those of a diagonal RIS. This result is expected because the low-complexity BD-RIS solution performs a projection onto the set of unitary and symmetric matrices, which is clearly suboptimal in terms of capacity. However, we have observed that the solution in \cite{MaoCL2024} provides a good initialization for the iterative algorithm proposed in this paper (typically better than a random initialization).

\begin{figure}[htb]
    \centering
\includegraphics[width=.5\textwidth]{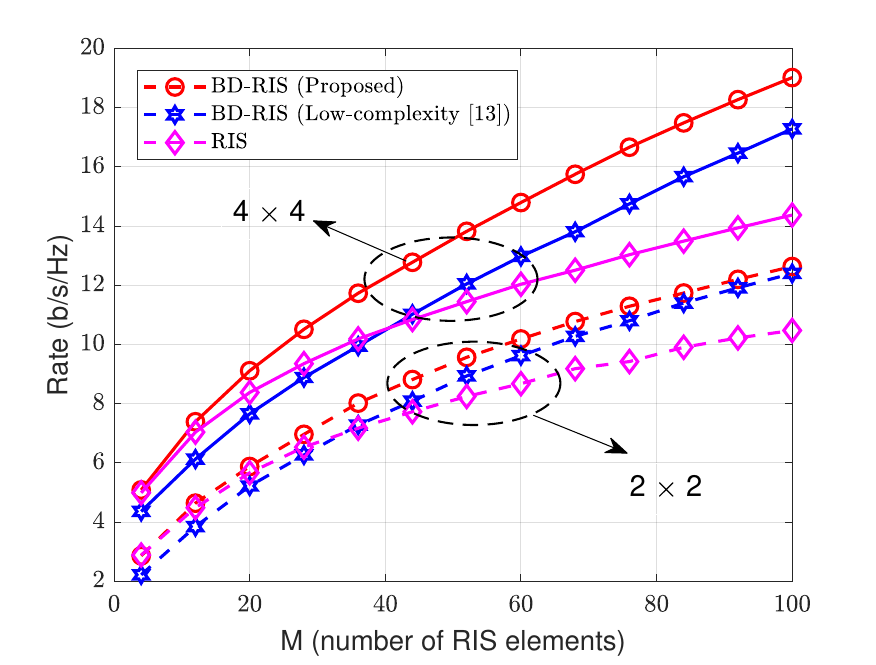}
     \caption{Achievable rates for an increasing number of BD-RIS elements $M$ for $2\times 2$ and $4\times 4$ MIMO channels.}     
     \label{fig:FigRateSPAWC}
\end{figure}

\paragraph{{\bf Results varying the number of BD-RIS elements}}
Fig. \ref{fig:FigRateSPAWC} shows the rates when the number of BD-RIS elements ranges from $M=10$ to $M=100$ when the RIS coordinates are (50,5,5) for $2\times 2$ and $4\times 4$ MIMO channels. For a $2\times2$ channel, the solution in \cite{MaoCL2024} is quite competitive, especially for higher $M$ values. As the number of streams of the MIMO system grows the differences between the proposed BD-RIS solution and the method in \cite{MaoCL2024} become more significant. 
%To further show this effect, Fig. \ref{fig:FigRateImprovementSPAWCN8} shows the relative improvement over a diagonal RIS of the solution proposed in this paper and the method in \cite{MaoCL2024} for an $8\times8$ system.

%\begin{figure}[htb]
%    \centering
%\includegraphics[width=.5\textwidth]{images/FigRateImprovementSPAWCN8_bis}
%     \caption{Relative rate improvement of a BD-RIS over a diagonal RIS for an increasing number of RIS elements for an $ 8\times 8$ MIMO system.}     \label{fig:FigRateImprovementSPAWCN8}
%\end{figure}

%\paragraph{{\bf Algorithm convergence}}
%Fig. \ref{fig:Convergence} shows the proposed algorithm's convergence curves analyzing the impact of initialization. The curves in red dashed line show the convergence of the proposed BDRIS algorithm with different random initializations. The iterations on the x-axis refer to the iterations of the outer loop alternating optimization (AO) process. Typically the algorithm converges in around 5 AO iterations. The dotted blue line shows the solution of the non-iterative algorithm in \cite{MaoCL2024} and the solid red line shows the convergence of the proposed algorithm with this solution as the starting point. As a comparison, we also show the convergence of the algorithm in \cite{ZhangCapacityJSAC2020} for diagonal RIS.

%\begin{figure}[htb]
%    \centering
%\includegraphics[width=.5\textwidth]{images/Convergence}
%     \caption{Convergence curves of the BDRIS and RIS iterative algorithms.}     
%     \label{fig:Convergence}
%\end{figure}

\begin{figure}[htb]
    \centering
\includegraphics[width=.5\textwidth]{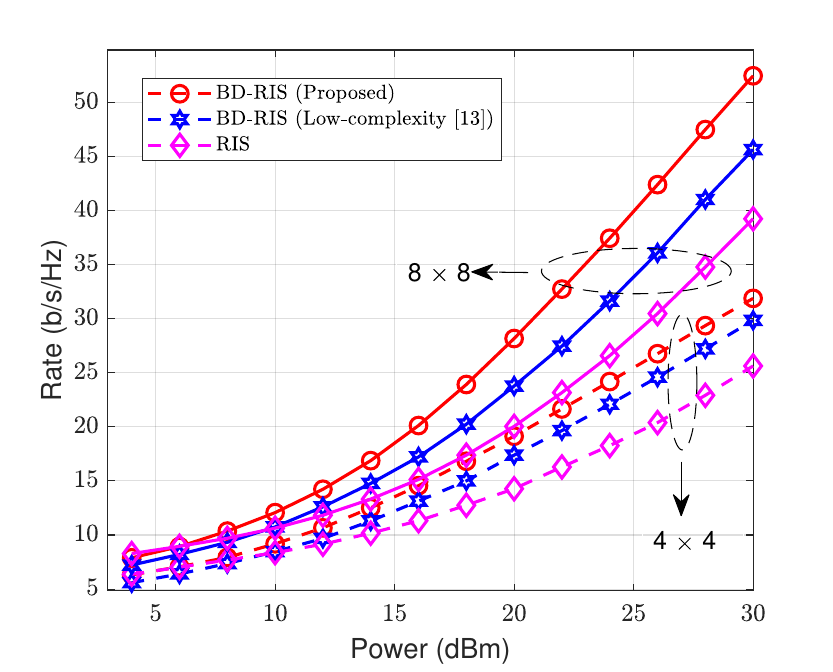}
     \caption{Achievable rate vs. transmit power for BD-RIS and RIS.}     
     \label{fig:FigSPAWCvsPower}
\end{figure}

\begin{figure}[htb]
    \centering
\includegraphics[width=.5\textwidth]{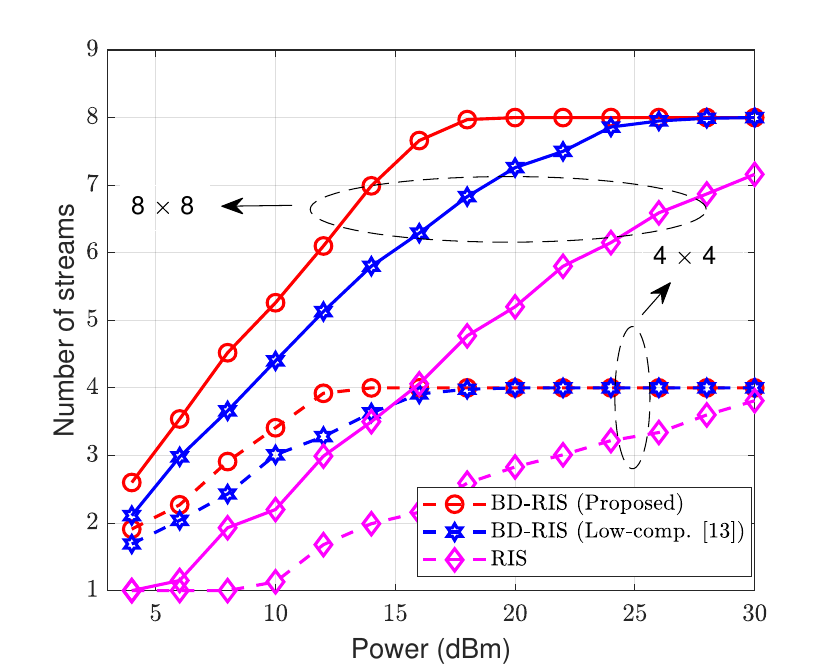}
     \caption{Number of active streams vs. transmit power for BD-RIS and RIS architectures.}     
     \label{fig:FigSPAWCvsStreams}
\end{figure}

\paragraph{{\bf Results varying the Tx power}}
Finally, Fig. \ref{fig:FigSPAWCvsPower} shows the achievable rates obtained by varying the input power between 4 and 30 dBms (which corresponds approximately to SNRs at the receiver between 15 and 35 dB) for $4 \times 4$ and $8 \times 8$ MIMO links assisted by a BD-RIS with $M=100$ elements. The improvements of the proposed algorithm are more appreciable for higher transmitting powers. Finally, Fig. \ref{fig:FigSPAWCvsStreams} shows the number of active streams vs. the Tx power. This figure suggests that the capacity improvements of BD-RIS over RIS can be attributed to the fact that BD-RIS creates channels that are better conditioned to support the transmission of a larger number of streams (for lower SNR).

%%%%%%%%%%%%%%%%%%%%%%%%%%%%%%%%%%%%%%%%%%%%%%%%%%%%%%%%%%%%%%%%%%%%%%%%%%%%%%%%%%%%%%%%%%%%%%%%%%%%%%%%%%%%%%%%%%%%%%%%%%%%%
%%%%%%%%%%%%%%%%%%%%%%%%%%%%%%%%%%%%%%%%%%%%%%%%%%%%%%%%%%%%%%%%%%%%%%%%%%%%%%%%%%%%%%%%%%%%%%%%%%%%%%%%%%%%%%%%%%%%%%%%%%%%%
%%%%%%%%%%%%%%%%%%%%%%%%%%%%%%%%%%%%%%%%%%%%%%%%%%%%%%%%%%%%%%%%%%%%%%%%%%%%%%%%%%%%%%%%%%%%%%%%%%%%%%%%%%%%%%%%%%%%%%%%%%%%%

\section{Conclusions}\label{sec:conclusions}
We have proposed an algorithm for maximizing the capacity of a MIMO link assisted by a fully connected BD-RIS. The main difficulty of the resulting optimization problem arises from the unitarity and symmetry constraints that the BD-RIS scattering matrix must satisfy. We have applied Takagi's factorization to overcome this difficulty, which allows us to derive an optimization algorithm in the manifold of unitary matrices. Compared to a diagonal RIS, the improvements in the achievable rates are more significant as the number of streams, the number of reflector elements, or the transmitted power increases. The proposed algorithm can be extended to group-connected BD-RIS architectures, as well as to multi-user MIMO networks. The development and analysis of these extensions remain as future work.

\section{Acknowledgment}
This work is supported by the European Commission’s Horizon Europe, Smart Networks and Services Joint Undertaking, research and innovation program under grant agreement 101139282, 6G-SENSES project. The work of I. Santamaria was also partly supported under grant PID2022-137099NB-C43 (MADDIE) funded by MICIU/AEI /10.13039/501100011033 and FEDER, UE.

%%%%%%%%%%%%%%%%%%%%%%%%%%%%%%%%%%%%%%%%%%%%%%%%%%%%%%%%%%%%%%%%%%%%%%%%%%%%%%%%%%%%%%%%%%%%%%%%%%%%%%%%%%%%%%%%%%%%%%%%%%%%%
%%%%%%%%%%%%%%%%%%%%%%%%%%%%%%%%%%%%%%%%%%%%%%%%%%%%%%%%%%%%%%%%%%%%%%%%%%%%%%%%%%%%%%%%%%%%%%%%%%%%%%%%%%%%%%%%%%%%%%%%%%%%%
%%%%%%%%%%%%%%%%%%%%%%%%%%%%%%%%%%%%%%%%%%%%%%%%%%%%%%%%%%%%%%%%%%%%%%%%%%%%%%%%%%%%%%%%%%%%%%%%%%%%%%%%%%%%%%%%%%%%%%%%%%%%%

%\balance
\bibliographystyle{ieeetr}
\bibliography{ref2}

\end{document}